\documentstyle[12pt]{article}

\begin{document}

\begin{center}
{\bf Some questions to experimental gravity.}

\vspace*{1.5cm} S.M.KOZYREV

Scientific center gravity wave studies ''Dulkyn''.

e-mail: Sergey@tnpko.ru
\end{center}

\vspace*{1.5cm}

{\bf Abstract}

The interpretations of solutions of Einstein field's equations led to the
prediction and the observation of physical phenomena which confirm the
important role of general relativity, as well as other relativistic theories
in physics. In this connection, the following questions are of interest and
importance: whether it is possible to solve the gauge problem and of its
physical significance, which one of the solutions gives the right
description of the observed values, how to clarify the physical meaning of
the coordinates which is unknown a priori, and why applying the same
physical requirements in different gauge fixing, we obtain different linear
approximations. We discuss some of the problems involved and point out
several open problems. The paper is written mainly for pedagogical purposes.

\vspace*{1.5cm}

We point out several problems associated with the general framework of
experimental gravity. Analyzing the static, vacuum solution of
scalar-tensor, vector-metric, f(R) as well as string-dilaton theories one
can find the following. When the energy-momentum tensor vanishes some of
solutions field equations of these theories amongst the others equivalent to
Schwarzachild solution in General Relativity \cite{a0}. The fact that this
result is not surprising can be seen as follows. First let us recall that in
empty space we can express the field equations of scalar tensor theories
\cite{2}

\begin{equation}
R_{\mu \nu }-\frac 12Rg_{\mu \nu }=\frac{\omega \left( \phi \right) }{\phi ^2%
}\left( \phi _{,\mu }\phi _{,\nu }-\frac 12g_{\mu \nu }g^{\alpha \beta }\phi
_{,\alpha }\phi _{,\beta }\right) +\frac 1\phi \left( \phi _{\mu \nu
}-g_{\mu \nu }\Box \phi \right) ,  \label{e01}
\end{equation}

\begin{equation}
\Box \phi +\frac 12\phi _{,\mu }\phi ^{,\mu }\frac d{d\phi }\ln \left( \frac{%
\omega \left( \phi \right) }\phi \right) +\frac 12\frac \phi {\omega \left(
\phi \right) }R=0.  \label{e3}
\end{equation}

in case of constant scalar potential $\phi $. Thus, it would be natural to
expect that the field equation (\ref{e01}) don't depend on $\phi $ and
equivalent to its in Einstein theory. Moreover, for Jordan, Brans Dicke
scalar-tensor theory the Hawking theorem states that the Schwarzschild
metric is the only spherically symmetric solution of the vacuum field
equations \cite{a2}. Obviously, there are the similar ways to obtain the
vacuum solution of vector metric theories too \cite{2}. One can write the
field equations

\begin{equation}
R_{\mu \nu }-\frac 12Rg_{\mu \nu }=\omega \stackrel{\left( \omega \right) }{%
\Theta }_{\mu \nu }+\eta \stackrel{\left( \omega \right) }{\Theta }_{\mu \nu
}+\varepsilon \stackrel{\left( \omega \right) }{\Theta }_{\mu \nu }+\tau
\stackrel{\left( \omega \right) }{\Theta }_{\mu \nu },  \label{e02}
\end{equation}

\begin{equation}
\varepsilon F_{;\nu }^{\mu \nu }+\frac 12\tau K_{;\nu }^{\mu \nu }-\frac
12K^\mu R-\frac 12\eta K^\nu R_\nu ^\mu =0.  \label{e03}
\end{equation}

where {\it K} vector field, {\it F}$_{\mu \nu }=K_{\nu ;\mu }-K_{\mu ;\nu },$%
and $\omega ,\eta ,\varepsilon ,\tau $ free parameters. Indeed, for the
absence of vector potential in empty space equations (\ref{e02}) become
identical to field equation of Einstein theory. {\it In this case no
celestial-mechanical experiments to reveal a difference between gravity
theories is not presented possible, since all Einstein's vacuum solutions
(Schwarzschild, Kerr, etc.) will satisfy these theories too.}

On the other hand in alternative relativistic theories inside the mater
gravitation is greatly distinguished from Einstein theory. Indeed, in the
stuff we have a specific physical situation. For example, scalar field in
Jordan, Brans - Dicke theory, inside the matter has characteristics like
gravitation permeability of material similar electromagnetic permeability of
material in Maxwell theories of electromagnetism \cite{a3}. Thus, in this
case the differences between relativistic theories should be experimentally
tested in substance. In this connection the dark matter problem may be
explained by variations of mater properties inside the galaxies and galaxies
clusters, with the same boundary condition.

It is commonly believed that general theory of relativity is a gauge theory
like any other (e.g., electromagnetism and Yang-Mills theory). However, from
the physical point of view it is radically different, just because of its
invariance under a group of diffeomorphisms acting on space-time itself,
instead of being invariant under the action of a local inner Lie group. In
this connection, in general theory of relativity we cannot rely from the
beginning on empirically validated, gauge-invariant dynamical equations for
the local fields, as it happens with electromagnetism, where Maxwell
equations can be written in terms of the gauge invariant potentials. The
gauge of General Relativity is the unphysical degree of freedom and we have
to fix the gauge to obtain physical results. {\it Moreover, in General
Relativity at present there is no clear understanding both of the gauge
problem and of its physical significance \cite{a4},\cite{a6}. Thus,
treatments of gauge are crucial in general relativity. Consequently one has
to find theoretical or experimental reasons to resolve this essential
ambiguity.}

According to the standard textbooks the Einstein equations do determine the
solution of a given physical problem up to four arbitrary functions. The
space-time geometry is constrained by the six equations, which place six
independent constraints on the ten components of the metric g$_{ij}$.

It is well known that in the gauge theories we may have different solutions
with the same symmetry in the base space. On the contrary, the gauge freedom
of Einstein's theory is such that the introduction of extra variables by
ruling out any background structure at the outset, it makes its physical
interpretation more deceptive and conceals at the same time the intrinsic
properties of point-events. The change of coordinates on the base space
induces automatically a nontrivial change of the frames on the fiber of
frames. Singular coordinate transformations may produce a change of the
gauge sector of the solution, because they may change the topology of the
frame bundle, adding new singular points and singular submanifolds, or
removing some of the existing ones. As a result, under singular gauge
transformation the solution of some initial physical problem will be
transformed onto a solution of a completely different problem.

A classical relativity theory is constructed inspired by the principle that
physical results of any theory must not depend on the choice of the
variables and, in particular, these results must be invariant under changes
of coordinates. However, it is obvious that the change of the interpretation
of the variables may change the formulation of the mathematical problem and
thus, the physical results, because we are using the variables according to
their meaning. Thus, applying the same physical requirements in different
variables, we arrive at different physical theories, because we are solving
field equations under different boundary conditions.

One of the largest concentrations of literature within the apparently simple
problem of General Relativity is that of solutions Einstein equations for
static perfect fluid sphere. However, the strong believe in the independence
of the relativity theory result on the choice of coordinates predisposes us
to a somewhat frivolous attitude towards the choice of the coordinates. For
the spherically symmetric configurations, the coordinates, which are
essentially different somewhere else, may be locally equivalent in the
vacuum. However, in other case all local relativity theory effects, like
gravitational redshift, perihelion shift, deflection of light rays,
time-delay of electromagnetic pulses, etc., will not have their standard
values \cite{a6}, and all the more so, it may be more essential for the
quantities, which depend on the precise form of the potential. The physical
and geometrical meaning of the coordinate {\it r} is not defined by
spherical symmetry of the problem and is unknown a priori \cite{a6a}, \cite
{a6e}. As a result, there are infinitely many static solutions of Einstein
equations of spherical symmetry, some of them are discovered at the early
stage of development of General Relativity, but up to now they are often
considered as equivalent representations of some unique solution. Some of
these solutions fall into different gauge classes, which describe physically
and geometrically different space-times \cite{a4}. In these cases, for
example, there is match more freedom in allowed relativistic stellar
configurations. One can use various coordinate conditions to choose {\it r}
and, in contrast with standard astrophysics, describe extreme objects with
arbitrary large mass, density and size \cite{a4}, \cite{a6s}.

As we have seen, the choice of the radial coordinate in relativistic
gravitational theories in a proper sense is essential for the description of
its gravitational field and needs a careful analysis. The different
coordinates in a given frame are equivalent only locally. Now the problem is
to clarify the physical meaning of the coordinate in relativistic
gravitational theories. The most important physical problem becomes to
interpretations for an experimental and observational probe of different
solutions of Einstein equations with spherical symmetry. Since there are the
physical and geometrical differences between the solutions of this type, one
must find a real way of distinguishing them experimentally. {\it Especially,
a problem of the present day is to answer the question about which one of
the spherically symmetric solutions gives the right description of the
observed values.}

In order to confront the predictions of a given gravity theory with
experiment in the solar system, it is necessary to compute its weak and
quasi-stationary fields or post-Newtonian approximation. This limit has been
computed for many metric theories of gravity and put in a standardized form
\cite{2}, which depends on a set of parameters that change from theory to
theory. Solar-system experiments allowed one to map out fairly completely
weak-field gravity at the first post-Newtonian approximation, i.e., to put
stringent numerical constraints on a large class of possible deviations from
General Relativity at order 1/c$^2$. Nordtvedt, Will and others were led to
provide rigorous underpinnings to the operational significance of various
theories, especially in solar system context, developing the parameterized
post Newtonian formalism (PPN)\ as a theoretical standard for expressing the
predictions of relativistic gravitational theories in terms which could be
directly related to experimental observations. However, for the case of
identity of vacuum solutions of General Relativity, scalar-tensor,
vector-metric, f(R) as well as string-dilaton we come to a conclusion that
PPN parameters in empty space in given relativistic theories are similar.
This means to lose any general prescription and to obtain corrections to the
PPN parameters in external and internal areas of material objects.
Consequently, the post-Newtonian approximations have to be handled very
carefully because the results could not be equivalent. The issue of the
correct Newtonian and post-Newtonian limit of such theories is still open.

In general, any relativistic theory of gravitation can yield corrections to
the Newton potential which, in the post-Newtonian formalism, could furnish
tests for the same theory \cite{2}. Howsoever, it will be outlined here that
the Newtonian limit cannot reproduce all the solutions of the relativistic
theories. This again emphasizes the fact that the structure of field
equations is deeply connected with the structure of spacetime. Another
apparent problem with the limit to General Relativity lies in the fact that
the scalar tensor theories does not tend to general relativity in the $%
\omega $ $\rightarrow $ $\infty $ limit. This feature is significant, for
example for Jordan, Brans Dicke theory, because the lower limit of $\omega >$%
~ 10$^4$ \cite{a2a} for the solar system measurements is fixed using the O$%
\left( \frac 1\omega \right) $ behavior in the standard PPN approximation.
However, recently it has been shown that for the Jordan, Brans Dicke theory,
which is the simplest scalar tensor theory the asymptotic behavior of $%
\omega $ is not represented by equation O$\left( \frac 1\omega \right) $ but
follows the relation O$\left( \frac 1{\sqrt{\omega }}\right) \cite{a2b}.\ $%
It is therefore important to study the situation more closely. If, following
Hawking theorem, we make the reasonable demand that the solution of
scalar-tensor theory field equations in empty space is the Schwarzschild
solution lead to free estimates of lower limit of $\omega .$

As is well known the largest effect for objects with slow velocities comes
from the g$_{00}$ component of the metric. Less obvious is the
interpretation of the g$_{ij}$ components which can be tested by the
measurement of spatial distances. Indeed space is not assumed to be flat.
One can infer that distances measured in the direction of the field gradient
behave differently as in other directions. Another interpretation assumes
that time is absolute, space is flat and absolute and gravitational field is
presented by two potentials: one scalar and one vector, which produces the
Coriolis force \cite{a9}, \cite{a10}. {\it Thus passageway to ''Newtonian''
limit does not lead to Newtonian theory.}

As matter of fact, the Newtonian gravity, after the pioneer article by
Cartan \cite{a11}, can be formulated as some sort of degenerate limit of
General Relativity \cite{a12}, \cite{a13}, \cite{a14}. In these
formulations, one has to deal with the fact that the metric becomes,
singular, and hence there no longer exists a Levi-Civita connection. In
further typical feature of this geometrical formulation is the fact that the
curved four dimensional affine structures can be separated out at a flat
affine structure and gravitational potential, by introducing the so-called
Cartan connection, this splitting, however, can not be done in unique way,
unless special boundary conditions are externally provided. The geometrical
four dimensional model of the Newtonian theory of gravitation shows that the
singular metric tensor plays the role of the gravitational tensor potential
determined to certain gauge transformation. The metrics of Newtonian
space-time is defined ambiguously for a given gravitational field. This
theory describes, besides Newtonian gravitational fields, also rotational
gravitational fields which are characterized by the fact that the Coriolis
type gravitational forces cannot be removed by any choice of rectangular
Cartrsian coordinate system \cite{a15}. {\it Consequently, this geometrical
formulation of the gravitational theories does not reproduce Newtonian
gravity but generalize those introducing corrections to the Newtonian
potential.}

The socalled linear approximation discussed in standard literature is often
utilizes explicitly the flat character of the Minkowski space-time manifold
and its formulation is essentially based on the existence of Cartesian
coordinates.

\begin{equation}
g_{\mu \nu }={\it \eta }_{\mu \nu }+h_{\mu \nu },\ \left| h_{\mu \nu
}\right| \ll 1.  \label{e1}
\end{equation}

where ${\it \eta }_{\mu \nu }$ is the Minkowski metric and $h_{\mu \nu }$
are small deviations which are treated only to first order wherever they
occur. Although this form of the metric restricts the metric to be nearly
Minkowskian one has a remaining freedom to choose a coordinate system. For
instance, to calculate gravitational waves as small perturbations in the
vacuum one usually uses the freedom of coordinate (or gauge) transformations
to impose the restriction harmonic gauge also known as wave coordinates on
the gravitational field thus fixing to a certain amount the coordinate
system. After imposing this special gauge conditions the Einstein equation
becomes a system of wave equations. The solution to this equation is
constructed as the disappearing on infinity retarded potential. However in
contrast with electrodynamics we can't discard advanced potential or its
combinations as a solution for gravitation field. {\it Consequently the
speed of gravity wave should not be fixed apriori and need the experimental
verification \cite{a71}.}

The metric of space-time were obtained by using the simplifying properties
of Einstein-Fock-de Donder gauge \cite{a7}

\begin{center}
\begin{eqnarray*}
g_{00} &=&1+2U, \\
g_{i0} &=&g_{0i}=4U_i, \\
g_{ik} &=&-\left( 1-2U\right) \delta _{ik}, \\
i &=&k=1,2,3.
\end{eqnarray*}
\end{center}

where {\it U} is a Newton's potential$.$

Another choice of gauge (Fermi coordinates) would lead to a different metric
in a weak field regime \cite{a8}.

\begin{eqnarray*}
g_{00} &=&-1+2\left( \phi \left( 0\right) -\phi \left( \overrightarrow{x}%
\right) \right) , \\
g_{0i} &=&0, \\
g_{ij} &=&\delta _{ij}-\gamma \underline{_{ij}}
\end{eqnarray*}

where

\begin{eqnarray*}
\gamma \underline{_{ij}} &=&2\overrightarrow{x}^2\left\{ v_1\left[ \phi
,_{_{ij}}\right] -v_2\left[ \phi ,_{_{ij}}\right] \right\} +2\delta
_{ij}\left\{ \phi \left( \overrightarrow{x}\right) -2v_o\left[ \phi \right]
+\phi \left( 0\right) \right\} \\
&&-2\left[ x^i\left\{ 2v_1\left[ \phi ,_{_j}\right] -v_0\left[ \phi
,_{_j}\right] \right\} +x^j\left\{ 2v_1\left[ \phi ,_i\right] -v_0\left[
\phi ,_i\right] \right\} \right] .
\end{eqnarray*}

In geometrical sense the different solutions in relativistic theories define
different linear approximation of pseudo-Riemannian space-time manifolds. A
subtle point in General Relativity formalism is that transitions from a
given physical solution to an essentially different one can be represented
as a change of coordinates. Thus, applying the same physical requirements in
different gauge fixing, we obtain different linear approximations because we
are solving field equations under different boundary conditions. One can see
that the gauge fixing problem in the above sense is essential for the
description of gravitational field and needs a careful analysis for the
adequate explanation of its linear approximations. Now the problem is to
clarify the physical meaning of the coordinates which is unknown a priori.


\begin{thebibliography}{99}
\bibitem{a0}  V.Bashkov, S. Kozyrev, Problems of high-energy physics and
field theory, 22, Protvino, (1991).

\bibitem{a2}  S. W. Hawking, Commun.Math. Phys.25, 167 (1972).

\bibitem{a3}  S. Kozyrev, Properties of the static, spherically symmetric
solutions in the Jordan, Brans-Dicke theory, gr-qc/0207039 (2002)

\bibitem{2}  C. M. Will, Theory and experiment in gravitational physics,
(2nd edn.), Cambridge University Press, Cambridge, (1993).

\bibitem{a2a}  B. Bertotti et al, Nature (London) 425, 374 (2003).

\bibitem{a2b}  A. Bhadra, K.K. Nandi, $\omega $ dependence of the scalar
field in Brans-Dicke theory, gr-qc/0409091 (2004).

\bibitem{a4}  P. P. Fiziev, Gravitational Field of Massive Point Particle in
General Relativity, gr-qc/0306088 (2003).

\bibitem{a6}  A. Gullstrand, Arkiv for matematik, astronomi och fysik,16, 8,
(1921).

\bibitem{a6a}  J.L.Synge, Relativity: The General Theory, North Holland
Publ. Comp., Amsterdam, 1960.

\bibitem{a6e}  A.S.Edington, The mathematical theory of relativity, 2nd. ed.
Cambridge, University Press, 1930.

\bibitem{a6s}  J. M. Aguirregabiria, Ll. Bel, Extreme objects with arbitrary
large mass, or density, and arbitrary size gr-qc/0105043, Gen. Rel. and
Grav. 33, 2049 (2001).

\bibitem{a7}  V. A. Fock, The Theory of Space, Time and Gravitation,
Pergamon, Oxford, 1964.

\bibitem{a71}  C.Moller, The Theory of Relativity, Oxford University Press,
Oxford, 1952.

\bibitem{a8}  Karl-Peter Marzlin, Fermi coordinates for weak gravitational
fields, gr-qc/9403044 (1994).

\bibitem{a9}  A. Einstein. In book: Mach Principle: From Newtons Bucket to
Quantum Gravity Eds.: J. Barbour, H. Pfister (Birkhauser, Berlin, 1995), p.
182

\bibitem{a11}  E. Cartan, Sur les varietes `a connexion affine et la theorie
de la relativite generalisee, Ann. de L'Ecole Normale. Sup. 40, 325, (1923).

\bibitem{a12}  A. Trautman, Comparison of newtonian and relativistic
theories of space-time, Perspectives in geometry and relativity (B.
Hoffmann, ed.), Indiana University Press, 1966.

\bibitem{a13}  W. Dixon, On the uniqueness of the newtonian theory as a
geometric theory of gravita- tion, Comm. Math. Phys. 45,no. 2, 167-182,
(1975).

\bibitem{a14}  P. Havas, Rev. Mod. Phys. 36, 938,\ (1964).

\bibitem{a15}  H. Keres, The gravitational tensor potential and Cartan's
four-dimensional representation of Newtonian thery, Izv. Ak. Nauk Est. SSR,
v. 25, 4, 349-358, (1976) (in Russian)

\bibitem{a10}  P. F. Browne, J. Phys. A: Math. Gen., 10 (1977).
\end{thebibliography}
\end{document}